# Solar neutron and muon detection on November 11, 2025: First simultaneous recovery of energy spectra.

*A. Chilingarian, B. Sargsyan, L. Kozliner, and T. Karapetyan*

A. Alikanyan National Lab (Yerevan Physics Institute)
Alikhanyan Brothers 2, Yerevan 36, Armenia, AM00036

**Abstract**

Ground Level Enhancement (GLE) events provide rare opportunities to study high-energy solar particle acceleration through direct detection of secondary radiation at ground level. On November 11, 2025, the Aragats Solar Neutron Telescope (ASNT) recorded a statistically significant increase in high-energy neutron and muon fluxes associated with an X5.1 flare and the subsequent Solar Energetic Proton (SEP) event. The event displayed a unique dual-peak profile: an initial hard component at 10:28 UT, followed by a softer yet still energetic peak at 10:45 UT. For the first time, we report simultaneous energy spectra of atmospheric neutrons and muons measured in the 10–600 MeV range at Aragats. Broken-power-law fits reveal a clear temporal evolution of acceleration conditions, evidenced by spectral indices declining with energy. These findings highlight the unique capabilities of the ASNT as an instrument for studying extreme solar particle acceleration.

**Plain Language Summary**

On November 11, 2025, a powerful storm of energetic particles from the Sun reached Earth. High-altitude detectors in Armenia measured fast neutrons and muons produced in Earth's atmosphere by these particles. The signal showed two separate bursts, only 11 minutes apart. By analyzing the particle energy distribution, we found that the first burst contained many more high-energy particles than the second. This reveals that the solar accelerator — possibly a shock wave driven by the solar eruption — weakened rapidly. These results help scientists understand where and how the most dangerous radiation from the Sun is created. Unique characteristics of the Aragats Solar Neutron Telescope allowed obtaining valuable information on neutron and muon energy spectra not available from other networks monitoring solar particle fluxes.

1. **Introduction**

Ground-Level Enhancements (GLEs) are rare episodes in which relativistic solar particles penetrate Earth's atmosphere and produce significant secondary radiation detectable by ground-based detectors (Shea & Smart, 2012). GLEs are the highest-energy manifestation of solar energetic proton (SEP) events, observed when relativistic ions accelerated in the vicinity of the Sun enter the terrestrial atmosphere and generate secondary radiation. More than 70 s have been observed since 1942, predominantly associated with major eruptive flares and CME-driven shocks (Bütikofer & Flückiger, 2015).

Energy spectra of muons and neutrons produced by solar proton interactions in the atmosphere serve as direct indicators of the acceleration environment in the SEP and flare regions (Reames, 2021). Their overall production spectrum index closely relates to the energy distribution of shock- or reconnection-accelerated ions. Therefore, precisely characterizing the spectral shapes offers rare insights into the time-dependent acceleration processes on the Sun.

At Mt. Aragats, the Advanced Solar Neutron Telescope (ASNT) detects energetic neutrons and muons and can differentiate between neutral and charged particles (Chilingarian et al., 2007). Its large size, thick scintillating spectrometer, and logarithmic amplitude-to-digital converter enable simultaneous measurement of muon and neutron energy spectra from MeV to sub-GeV levels. Past studies have reconstructed GLE proton spectra using neutron monitor yield functions (e.g., Tylka & Dietrich, 2009). While useful, these methods are inherently indirect, relying on model-dependent inversion from numerous monitors with varying geomagnetic cutoff rigidities and atmospheric depths. Uncertainties in yield functions, the evolution of anisotropy, and transport conditions constrain accuracy at the highest energies, where the physics of acceleration is most restrictive.

In contrast, the ASNT spectrometer directly measures the secondary neutrons and muons produced by energetic hadronic cascades initiated by >10 GeV solar protons. Another key advantage is that ASNT detectors simultaneously track changes in the muon flux. Since muons mainly come from charged pion decay ($\pi\pm$), their spectral shape offers an independent way to analyze the primary proton spectral slope and helps differentiate between deep atmospheric cascades that produce muons at the detector and higher-altitude hadronic interactions that only generate neutrons. Combining neutron and muon spectroscopy enables consistent modeling of pion production, cascade attenuation, and transport within the geomagnetic field.

ASNT muon observations provide an independent diagnostic of pion production depth and angular distribution in the atmosphere, allowing estimation of anisotropy and event-integrated directional transport. The neutron-to-muon ratio further constrains where secondary interactions occurred and how sharply peaked the initial proton beam was when entering the magnetosphere. In this way, the combined dataset allows the disentangling of source physics and heliospheric transport processes, which previously were largely entangled in GLE interpretation.

This joint spectrum gives access to the most hazardous part of the distribution: multi-GeV protons that dominate radiation dose at aviation altitudes and in near-Earth space (Spence et al., 2013). The ASNT spectral peaks, together with satellite time profiles, allow one to distinguish an early, harder phase, likely associated with flare-site reconnection, from a later, softer phase, more characteristic of CME-driven shock acceleration (Gopalswamy et al., 2012).

In this paper, we present, for the first time, simultaneously measured energy spectra of GLE muons and neutrons and discuss the physical consequences arising from this new type of information.

   2. **ASNT neutron/muon spectrometer**

The Aragats Solar Neutron Telescope (ASNT) is designed to measure direct neutron fluxes from energetic solar flares and secondary muons and neutrons produced in Earth's atmosphere (Muraki et al., 2007). ASNT consists of four identical modules, as shown in Fig. 1. Each module includes forty scintillator slabs measuring 50 x 50 x 5 cm³, stacked vertically on a lower horizontal plastic scintillator slab of 100 x 100 x 10 cm³. Four scintillators, each 100 x 100 x 5 cm³, are positioned above the main assembly to discriminate neutral particles by "vetoing" charged.

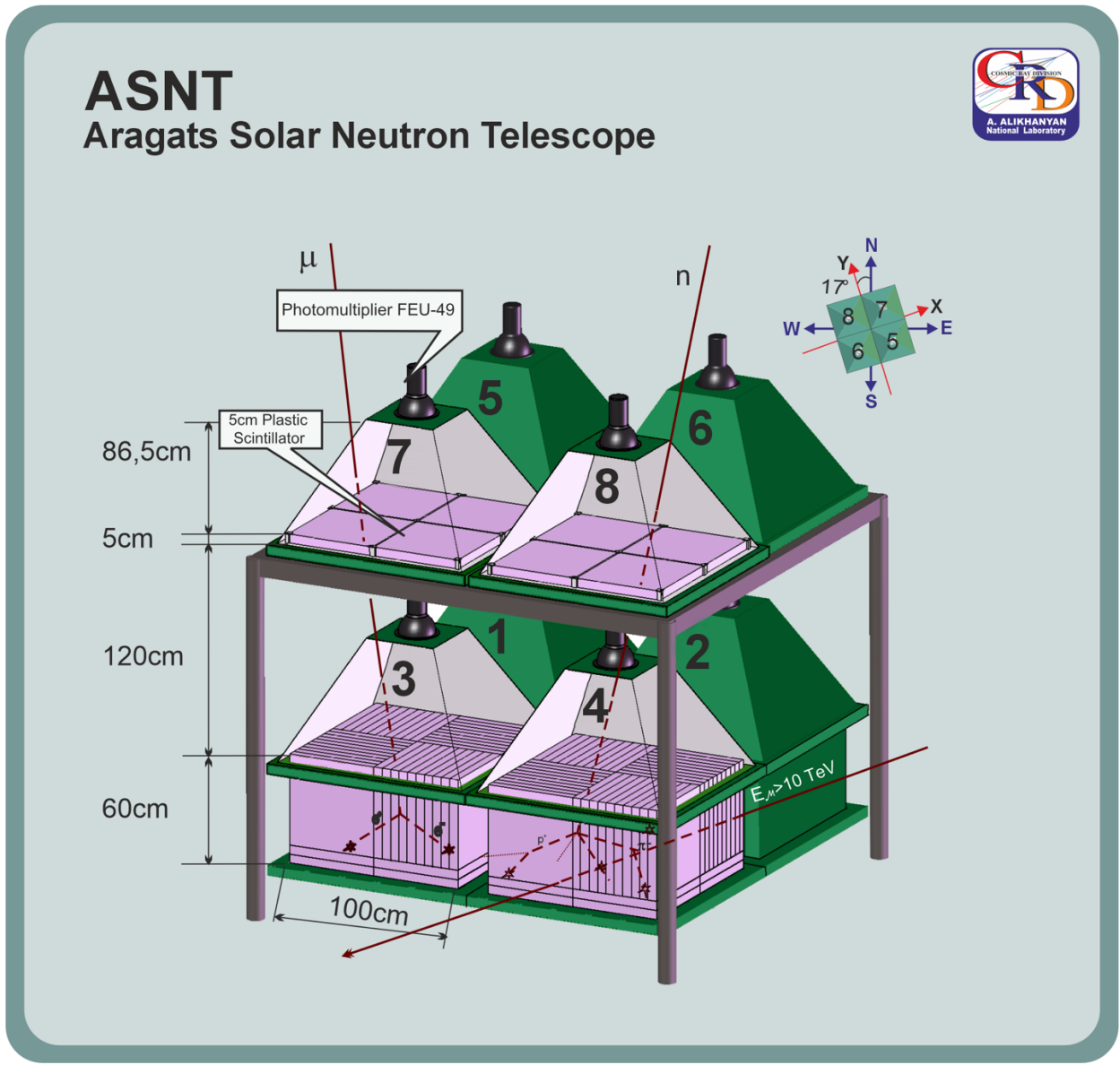

**Figure 1. Assembly of ASNT with the enumeration of 8 scintillators and orientation of detector axes relative to the North direction.**

The main ASNT trigger reads and stores the analog signals from all 8 channels if at least one channel reports a signal above threshold. The mean trigger frequency is ≈4 kHz, due to the ambient population of different cosmic ray (CR) species. Additional flux from GLEs usually does not exceed 5-10%. However, during thunderstorm ground enhancements (TGEs; Chilingarian et al., 2010), the flux of gamma rays and electrons can be up to 5 times the background level.

The available information from ASNT includes:

1. A 10-second time series of count rates for all 8 channels of ASNT, with the integration time of the scintillators adjustable from 1 to 60 seconds.

2. Count rates of particles arriving from different incident directions: 16 possible coincidences between 4 upper and 4 lower scintillators.

3. Count rates of 8 special coincidences, such as one signal from the upper scintillators and one from the lower, or no signals in the upper with more than one in the lower, etc. All coincidences are recorded if signals occur within a 1-microsecond window.

4. Estimates of the variances of count rates for each ASNT channel, calculated from 12 five-second counts—meaning that in a minute, 12 sets of counts (each with a 5-second integration) are stored, and then means and variances are computed from these values.

5. An 8 x 8 correlation matrix of ASNT channels, calculated from five-second count rates over one minute; the same stored five-second series is used each minute to monitor potential cross-talk between channels.

6. Each minute, or after July 2012, every 20 seconds, histograms of energy releases in all 8 channels of ASNT are recorded.

7. Similar data as in point 6, but only for particles not registered in the upper layer (veto on charged particles to select samples enriched by neutral particles).

8. Special triggers permit storing the traversal of horizontal muons within the acceptance ranges of 85-90° and 89.5-90°.

In this paper, we use items (4), (6), and (7) to reconstruct neutron and muon spectra during GLE 77. Raw histograms from pre- and peak-intervals were constructed by subtracting the background histograms measured before GLE:

$$N_{\text{excess}}(E) = N_{\text{signal}}(E) - N_{\text{background}}(E)$$

Statistical uncertainties are Poisson-based:

$$\sigma(E) = \sqrt{N_{\text{signal}} + N_{\text{background}}}$$

## 2. Results: Time Profiles and Energy Spectra of GLE 77

The GLE 77 event on November 11, 2025, followed an X5.1 solar flare and an exceptionally fast coronal mass ejection (>1800 km/s), leading to a prompt relativistic arrival and strong particle anisotropy. Neutron monitor arrays worldwide detected the onset around 10:12–10:14 UT, with a peak that varied significantly depending on station cutoff rigidity. GLE 77 is notable as a dual-peak event separated by roughly 10 minutes, featuring hard time profiles in the SEP production range. Measurements at Mt. Aragats, supported by a favorable line-of-sight geometry (Sun near zenith over Aragats), showed significant variations in neutron and muon yields aligned with rapid changes in acceleration efficiency at the Sun. GLE77 ASNT observations uniquely provide energy spectra up to about 600 MeV, filling an important observational gap globally. This direct spectral evidence is very rare — ASNT measurements offer a key reference point between satellite-observable energies and the NM-dominated regime.

Figure 2 shows a 1-minute time series of count rates of ASNT neutron and muon channels for particles depositing more than 40 MeV in the 60 cm scintillators. Baselines and variances were measured during a quiet period before enhancement, and both curves are plotted in units of sigma.

The neutron curve displays two clearly separated enhancements. The first peak begins at around 10:27 UT, reaches a maximum of approximately 11 sigma near 10:30–10:31 UT, and then diminishes over several minutes. A second, broader neutron enhancement starts near 10:45 UT, with a peak amplitude of about 7 sigma. Between these two peaks, there is a genuine 10-minute gap during which the ASNT rate remains close to background levels. This pattern is difficult to explain through gradual pitch-angle diffusion of a single injection and instead suggests two separate episodes of high-energy proton release along the same magnetic field line.

The muon time series above 40 MeV displays the same two peaks, but with systematically lower amplitudes—about 4–5 sigma in the first peak and 3–4 sigma in the second. The onset times of the muon enhancements align with those of neutrons within the one-minute binning, indicating that the highest-energy primaries responsible for both neutral and charged secondaries arrive nearly simultaneously. The smaller relative amplitudes in muons already suggest that, compared to classic events such as GLE 69, GLE 77 produced fewer protons above several tens of GeV and/or that the proton beam was less tightly beamed along the local field line.

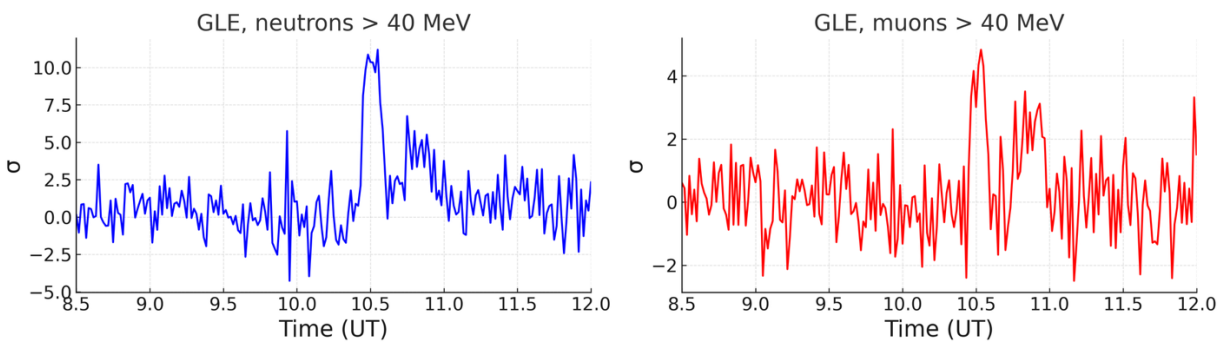

**Figure 2. GLE 77 time profiles measured by ASNT. Left: ASNT neutron channel, events with deposited energy > 40 MeV, expressed in σ units. Right: summed muon channel with the same energy threshold. Vertical structure of the first neutron peak (10:27–10:35 UT, max ≈11σ) and the broader second peak (10:45–10:54 UT, 6–7σ) is clearly visible; muons**

**also show a pronounced response. This behaviour already suggests that the later injection is softer and has fewer >10 GeV primaries.**

Energy spectra were reconstructed for both enhancements by converting the deposited energy in the 60 cm plastic scintillator of the ASNT and in the thick muon detector into integral spectra N(>E). The raw histograms were first background-subtracted (signal minus pre-event background), then unfolded using the energy-dependent registration efficiency obtained from GEANT4 simulations of the ASNT response. The resulting secondary neutron and muon spectra above Aragats were fitted in three energy bands (40–120, 120–260, and 260–600 MeV) with power laws of the form N(>E) ∝ $E^{-\gamma}$, Fig.3. We show energy spectra based on count rate only without normalization to detector area and unit time.

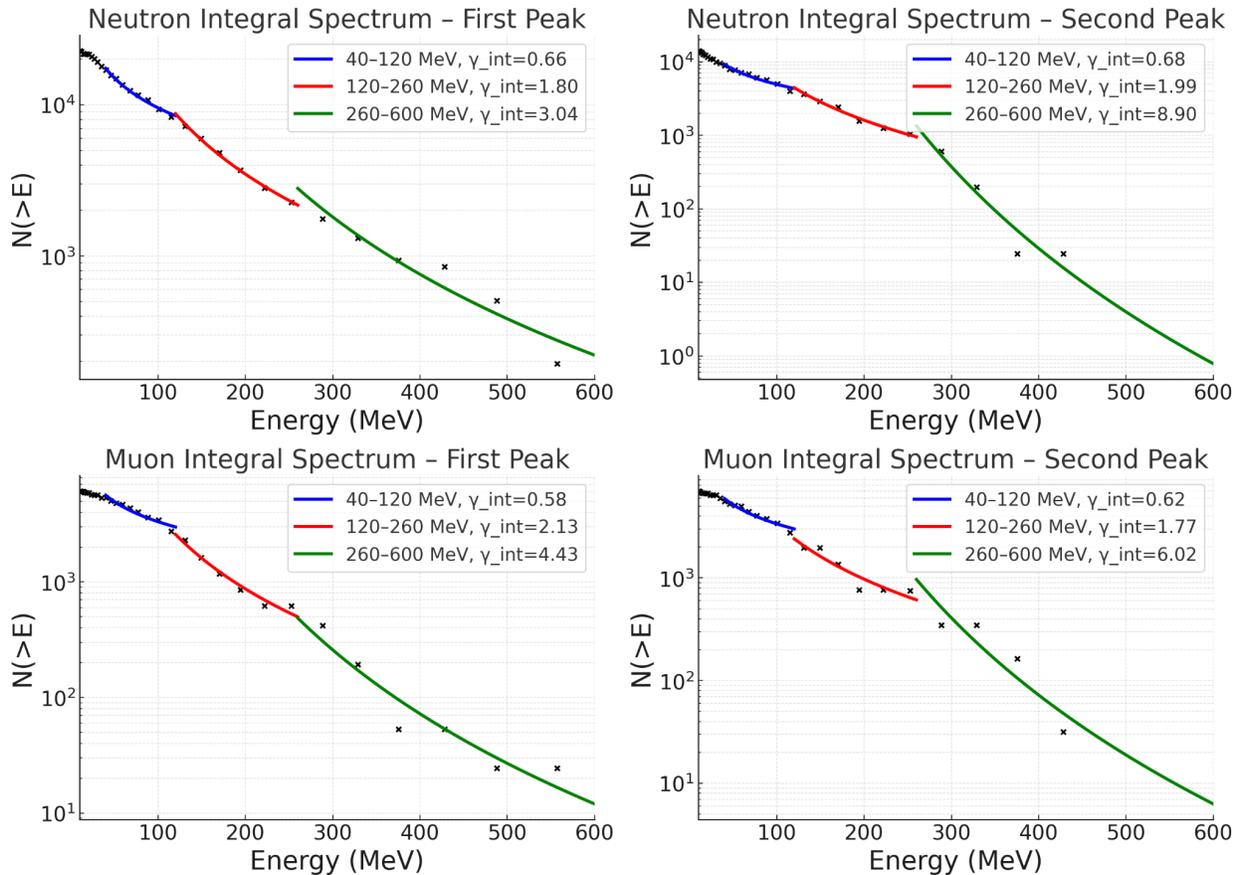

**Figure 3. Integral energy spectra of neutrons and muons during GLE 77. Top: unfolded neutron spectra for the first (left) and second (right) peaks. Bottom: muon spectra for the first (left) and second (right) peaks. Black symbols give the integral N(>E) points; coloured lines show power-law fits in three bands (40–120 MeV, blue; 120–260 MeV, red; 260–600 MeV, green). Both neutrons and muons exhibit hard spectra with γ ≈ 0.6–0.7 in the 40–120 MeV range, followed by progressive steepening above ≈120 MeV. The second peak is systematically softer than the first in both channels, especially in the 260–600 MeV band.**

The high-energy fits for the second peak are based on a few points with large errors, so that index value of 8.9 should be interpreted as "very steep" rather than a precise exponent.

Table 1 summarises the integral spectral indices γ (with 1σ errors) for neutrons and muons in the three energy bands for both peaks of GLE 77.

**Table 1. Integral power-law indices γ_int for neutrons and muons in three energy bands for the first and second peaks of GLE 77.**

| Component | Band (MeV) | γ ± error |
|---|---|---|
| 1 Neutrons | 40–120 | 0.66 ± 0.03 |
| 1 Neutrons | 120–260 | 1.80 ± 0.06 |
| 1 Neutrons | 260–600 | 3.04 ± 0.50 |
| 2 Neutrons | 40–120 | 0.68 ± 0.06 |
| 2 Neutrons | 120–260 | 1.99 ± 0.13 |
| 2 Neutrons | 260–600 | 8.90 ± 2.12 |
| 1 Muons | 40–120 | 0.58 ± 0.06 |
| 1 Muons | 120–260 | 2.13 ± 0.21 |
| 1 Muons | 260–600 | 4.43 ± 0.73 |
| 2 Muons | 40–120 | 0.62 ± 0.06 |
| 2 Muons | 120–260 | 1.77 ± 0.35 |
| 2 Muons | 260–600 | 6.02 ± 1.98 |

The nearly identical low-energy indices for neutrons and muons in the 40–120 MeV band ($\gamma \simeq 0.6$–$0.7$) indicate that this range is dominated by relatively fresh secondaries created in the atmosphere by primaries just above the local geomagnetic cut-off rigidity ($R_c \simeq 7.1$ GV). Above $\simeq 120$ MeV, the spectra soften: neutrons steepen to $\gamma_{int} \simeq 1.8$–$2.0$, while muons steepen even more strongly, reaching $\gamma_{int} \simeq 4$–$6$ in the 260–600 MeV band. This behaviour reflects the combined effects of hadronic production, decay kinematics, and energy losses of muons in the atmosphere and detector: only the highest-energy primaries can produce muons depositing 500–600 MeV in the 60 cm scintillator.

In both channels, their indices in the 260–600 MeV band are flatter than for the second peak, and the neutron and muon time series show larger statistical significance σ. The ≃10-minute gap between the maximum of the first and second neutron peaks, combined with the weaker and softer muon response during the second enhancement, suggests that the acceleration site (flare region and/or the early CME-driven shock) injected a harder proton spectrum during the first episode. By the time of the second peak, the proton distribution was already depleted in particles above ≃10 GeV, so that neutron production continued while the muon signal became marginal.

The neutron-to-muon ratio thus offers a qualitative constraint on the depth and angular distribution of the cascades. The fact that the first neutron peak is stronger than the second, while the relative muon amplitudes change less, indicates that the initial injection produced a more intense and somewhat harder proton beam, capable of initiating more high-altitude cascades. During the second peak, the overall intensity drops and the proton spectrum softens, but the persistence of a muon signal up to approximately 500–600 MeV deposited energy still requires primaries of at least several tens of GeV. There is no evidence of a significant broadening in the proton angular distribution between the two peaks: the neutron and muon time profiles remain sharp, and their ratio shows no dramatic evolution.

Taken together, the two-peak neutron and muon data suggest the following scenario: a first, relatively hard and well-collimated proton beam is injected near the time of the X5.1 flare maximum and reaches Earth with modest pitch-angle scattering, producing the stronger neutron and muon peak. A second injection occurs after approximately 10 minutes with lower intensity and a softer high-energy tail, resulting in weaker neutron and muon responses and steeper spectra above roughly 260 MeV. In this model, the neutron–to–muon ratio and its evolution mainly reflect changes in the maximum proton energy and the altitude distribution of pion production, while the close timing of the two channels constrains the amount of interplanetary scattering experienced by the highest-energy protons.

The detection of statistically significant muon signals up to deposited energies of 500–600 MeV implies parent proton energies of at least several tens of GeV when realistic hadronic interaction chains and propagation are taken into account. This is fully consistent with earlier Aragats studies of GLE 69, which demonstrated solar-proton spectra extending well beyond 20 GeV using high-energy muon observations (Bostanjyan et al., Chilingarian, 2009). GLE 77 thus confirms that even in solar cycle 25 the Sun can still accelerate particles into the multi-GeV domain, although the very steep indices of the 260–600 MeV muon band indicate a smaller relative contribution of >10–20 GeV protons than in the extreme 20 January 2005 event.

## 4. Discussion and conclusions

After 20 events of significant GLE absence, GLE 77 observations by the ASNT spectrometer naturally continued our studies of GLE 69, adding several new elements.

1. We obtain, for the first time, time-resolved neutron and muon spectra for two clearly separated peaks of a single GLE in solar cycle 25, using the same detector complex. The spectra are derived at the detector level (secondary neutrons and muons) using a well-calibrated response of the ASNT and associated scintillators, and then interpreted in terms of primary protons above the Aragats cut-off.
2. We demonstrate that both peaks of GLE 77 produce measurable high-energy muon enhancements, confirming that the accelerator in this event is capable of reaching at least the 10–20 GeV range in proton energy, similar to the strong GLEs of cycle 23 but at noticeably lower intensity. This again emphasizes the unique role of the thick-scintillator muon channel: neutron monitors alone have little sensitivity to such energies without additional assumptions on transport and anisotropy, whereas the muon signal directly tracks the high-energy tail of the cascade.
3. The combined neutron–muon analysis suggests that in GLE 77 the effective proton spectrum above ~10 GeV is softer than in the extreme 20 January 2005 event, as indicated by the steep muon indices and the comparatively modest muon excess We do not attempt a full forward-modelling comparison here; however, the steep muon indices and the modest muon significance, compared with the large neutron enhancements, both point to a relatively low contribution of ultra-relativistic protons. This qualitative picture is consistent with broader studies that generally find weaker high-energy proton acceleration in recent cycles than in the most extreme events of cycle 23 (Asvestari et al. 2017).
4. The clear two-peak structure, with the second peak softer and weaker in both neutrons and muons, demonstrates that the acceleration region evolves on time-scales of only a few tens of minutes. The early, harder peak is likely associated with the most efficient phase of flare and/or shock acceleration, close to the time of the X5.1 flare maximum and the onset of the >500 MeV proton flux at GOES. The later peak then reflects either a decaying flare arcade or a shock front propagating into a less favourable environment for accelerating protons to tens of GeV.
5. Solar-cycle context and evolution of acceleration efficiency.
Recent comparisons of solar energetic particle events across cycles 23 and 24–25 show that cycle 23 contained a larger fraction of extremely hard, very-high-energy events, while later cycles tend to produce softer GLE spectra with lower maximum energies. The combined neutron–muon analysis of GLE 77 fits naturally into this picture. The hard low-energy indices and the presence of a clear muon signal demonstrate that the Sun in cycle 25 still accelerates protons to several GeV. However, the steep muon spectra and the strong softening above a few hundred MeV indicate a lower $E_{max}$ than in GLE 69. The two-peak structure, with the second peak softer than the first, also agrees with models in which early acceleration near the flare and CME-launch region produces the highest-energy particles, while later phases are dominated by weaker or more extended acceleration at the CME-driven shock.
6. Methodological advance at Aragats.
Finally, the present work demonstrates a methodological advance for ASEC: the same detector complex (ASNT plus thick plastic scintillators) can now be used to recover energy spectra of both neutral (neutrons) and charged (muons) secondaries for solar events, in addition to its well-established role in thunderstorm ground enhancement

7. (TGE) studies (Chilingarian et al., 2010; 2024). The unfolding procedures developed for TGE spectra recovering have been successfully applied to a GLE induced spectra. This provides a powerful tool for future cycle-25 and cycle-26 events: whenever a significant solar neutron signal is accompanied by a measurable muon excess at Aragats, the combined spectral reconstruction will yield tight constraints on both the shape and the cutoff of the solar-proton spectrum.

In summary, GLE 77 confirms that the Sun is still capable of producing multi-GeV protons in solar cycle 25, but the reconstructed neutron and muon spectra indicate a smaller contribution of >10 GeV particles compared to the 20 January 2005 "benchmark" event. The key new element here is the simultaneous use of unfolded neutron spectra and thick-scintillator muon spectra from the same detector complex, which strongly constrains both the shape and the cutoff of the solar-proton spectrum. By measuring the neutron and muon yield above adjustable energy thresholds (40–260 MeV), we directly access the upper tail of the incident solar proton spectrum — a region that satellite detectors typically cannot constrain with sufficient statistics.

Neutron monitor reconstructions are most reliable below ~10 GeV; extending fits to higher energies becomes increasingly model-dependent. In contrast, at Aragats, the geomagnetic cutoff rigidity of 7.1 GV blocks most lower-energy primaries and influences the selected interactions of high-energy SEP particles. Therefore, obtaining neutron and muon spectra at ground level provides a crucial complement to satellite spectrometers (Ajello et al., 2021) and worldwide NM data (Mishev & Usoskin, 2020) extending spectral reconstruction into the multi-GeV region of parent solar protons.

The dual-component SEP behavior has been observed by spaceborne particle detectors during several past GLEs, but without direct, high-energy ground support. ASNT now provides that missing constraint. Before this work, no ground-based experiment had reported neutron and muon energy spectra from a GLE above 40 MeV. Atmospheric neutron spectroscopy during GLEs had been limited to integral NM rates and satellite-derived proton fits. Spaceborne instruments such as GOES directly measure the primary proton population, but their upper channels do not sufficiently constrain the relativistic tail of the spectrum, especially during short, impulsive phases of GLEs (Onsager et al., 1996). ASNT data are the only direct measurements of GLE-induced atmospheric neutrons and muons in the 40–600 MeV range, offering insight into the relativistic cutoff region of SEP acceleration near the Sun. The current results fill the observational gap between spaceborne SEP measurements—direct proton data—which are limited to 500 MeV. Combining satellite proton flux measurements with ground-based neutron and muon observations enables the reconstruction of the solar proton spectrum across a uniquely broad energy range. Satellite and ground-level datasets are inherently complementary: the former define the low-energy turnover and initial spectral slope, while the latter determine the high-energy continuation and potential spectral hardening.

Perhaps most importantly, ground-level muon and neutron observations reveal the portion of the SEP spectrum that controls radiation hazards for near-Earth space, aviation, and high-altitude technology. Protons above several GeV disproportionately contribute to dose rates at aircraft altitudes and can only be evaluated using high-energy secondary measurements. In this regard, ASNT results extend quantitative GLE risk assessment into a regime where neutron monitors provide limited information and satellite instruments are sensitivity-constrained. GLE 77 is a historically strong event providing a benchmark for space weather operations; its validated spectra help calibrate predictive tools for future radiation storms.